\documentclass[
    aps,
    prl,
    reprint,
    superscriptaddress
]{revtex4-2}

\usepackage{graphicx}
\usepackage{bm}
\usepackage{dcolumn}
\usepackage{amssymb}
\usepackage{amsmath}
\usepackage{parskip}
\usepackage{array,multirow}
\usepackage{multirow}
\usepackage{dcolumn}
\usepackage{afterpage}
\usepackage{rotating}
\usepackage{hyperref}
\usepackage{xcolor}
\usepackage{titlesec}
\usepackage{float}
\usepackage{ifxetex,ifluatex}
\usepackage{xcolor}
\usepackage[english]{babel}
\usepackage[sort&compress]{natbib}
\usepackage{xfrac}
\usepackage{graphicx}
\usepackage{threeparttable}
\usepackage{mathpazo}
\usepackage{textcomp}
\usepackage{nicefrac} 
\usepackage[perpage]{footmisc} 

\begin{document}

\title{Crystallographic imperfections and exotic superconductivity of UBe$_{13}$}

\author{Andreas Leithe-Jasper}
\thanks{Andreas Leithe-Jasper, Markus König, and Yusei Shimizu contributed equally to this work}
\affiliation{Max Planck Institute for Chemical Physics of Solids, 01187 Dresden, Germany} 
\author{Markus K\"onig}
\affiliation{Max Planck Institute for Chemical Physics of Solids, 01187 Dresden, Germany}
\author{Yusei Shimizu}
\affiliation{Max Planck Institute for Chemical Physics of Solids, 01187 Dresden, Germany}
\affiliation{  Institute for Solid State Physics, The University of Tokyo, 2778581 Kashiwa, Japan}
\author{Konstantin Semeniuk}
\affiliation{Max Planck Institute for Chemical Physics of Solids, 01187 Dresden, Germany}
\author{Primo\v{z} Ko\v{z}elj}
\affiliation{Max Planck Institute for Chemical Physics of Solids, 01187 Dresden, Germany}
\affiliation{Faculty of Mathematics and Physics, University of Ljubljana, Jadranska 19, SI-1000 Ljubljana, Slovenia}
%\affiliation{Jožef Stefan Institute, Jamova cesta 39, SI-1000 Ljubljana, Slovenia}
\affiliation{Jo\v{z}ef Stefan Institute, Jamova cesta 39, SI-1000 Ljubljana, Slovenia}
\author{Mitja Krnel}
\affiliation{Max Planck Institute for Chemical Physics of Solids, 01187 Dresden, Germany}
\author{Paul Simon}
\author{Wilder Carrillo-Cabrera}
\affiliation{Max Planck Institute for Chemical Physics of Solids, 01187 Dresden, Germany}
\author{Nazar Zaremba}
\affiliation{Max Planck Institute for Chemical Physics of Solids, 01187 Dresden, Germany}
\author{Marcel Naumann}
\affiliation{Max Planck Institute for Chemical Physics of Solids, 01187 Dresden, Germany}
\author{Ulrich Burkhardt}
\affiliation{Max Planck Institute for Chemical Physics of Solids, 01187 Dresden, Germany}
\author{Thomas Doert}
\affiliation{Faculty of Chemistry and Food Chemistry, Dresden University of Technology, 01062 Dresden, Germany}
\author{\\ Yurii Prots}
\affiliation{Max Planck Institute for Chemical Physics of Solids, 01187 Dresden, Germany}
\author{Alfred Amon}
\affiliation{Lawrence Livermore National Laboratory, 7000 East Ave, Livermore, CA 94550, USA}
\author{Elena Gati}
\affiliation{Max Planck Institute for Chemical Physics of Solids, 01187 Dresden, Germany}
\affiliation{Goethe University Frankfurt, 60438 Frankfurt a.M., Germany}
\author{Matthias Vojta}
\affiliation{Institute of Theoretical Physics, Dresden University of Technology, 01062 Dresden, Germany}
\author{Yuri Grin}
\affiliation{Max Planck Institute for Chemical Physics of Solids, 01187 Dresden, Germany}
\author{Elena Hassinger}
%\affiliation{Institute for Quantum Materials and Technologies, Karlsruhe Institute of Technology, Kaiserstraße 12, 76131 Karlsruhe, Germany}
\affiliation{Institute for Quantum Materials and Technologies, Karlsruhe Institute of Technology, Kaiserstra\ss e 12, 76131 Karlsruhe, Germany}
\affiliation{Max Planck Institute for Chemical Physics of Solids, 01187 Dresden, Germany}
\author{Eteri Svanidze}
\email{svanidze@cpfs.mpg.de}
\affiliation{Max Planck Institute for Chemical Physics of Solids, 01187 Dresden, Germany}
\thanks{Andreas Leithe-Jasper, Markus König, and Yusei Shimizu contributed equally to this work}

\date{\today}

\begin{abstract} 
\noindent The symmetry of the superconducting gap is related to the symmetry of the crystal structure. In unconventional superconductors, big changes of the critical temperature, critical field or even the gap structure can happen even for small perturbations of the lattice. In this letter, we use microstructuring to study aluminium-free crystals of UBe$_{13}$ which are expected to be closer to "perfect" material than previously studied aluminium-grown single crystals. We compare the effect of minuscule imperfections on the value of the critical temperature and critical field, which, in the case of UBe$_{13}$, has drastic effects, supporting its unconventional nature. We conjecture that this likely arises from the oxidation state of uranium, which is sensitive to its crystal environment. Our findings suggests that uranium-based materials provide not only an excellent reservoir of new unconventional phenomena, but also a way to identify the gap symmetry of unconventional superconductors.

%We conclude that both $A_{1u}$ and $E_{u}$ triplet superconducting states coexist in UBe$_{13}$, reinforcing its spin-triplet nature. 

\end{abstract}

\maketitle

\vspace{-20pt}
\section{Introduction}

Unconventional superconductivity, $i.e.$ superconductivity with strongly momentum-dependent pairing often driven by a non-phonon mechanism, is observed in many material classes -- from high-temperature cuprates and pnictides to low-temperature heavy-fermion materials \cite{Stewart2017, Zhou2021}. The response to disorder/imperfections can provide a lot of information on the underlying superconducting order parameter -- in particular, this is a common indication of whether or not a given superconductor is unconventional. Unconventional superconductors include several uranium-based compounds -- in fact,  among nearly two dozen of known uranium-based superconductors, nearly half are currently classified as unconventional, with several spin-triplet candidates \cite{Huxley2001, Ran2019a, Tsutsumi2012, Stewart2019, Aoki2019}. 

\begin{figure*}
        \centering
    \includegraphics[width=\linewidth]{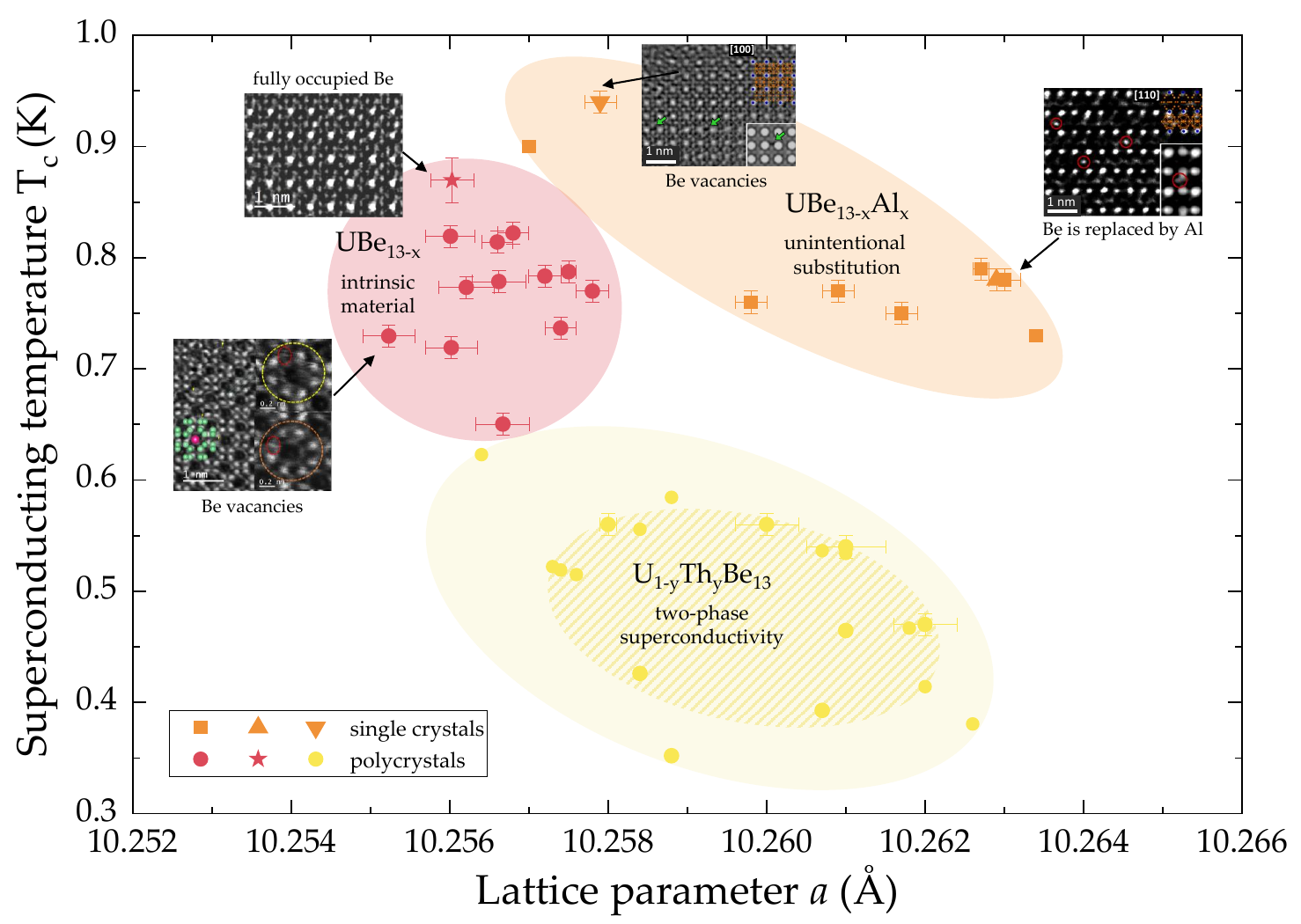}
    \caption{Interplay of chemistry and physics in U$_{1-y}$Th$_y$Be$_{13-x}$Al$_x$: superconductivity as a function of structural disorder, as reflected by the value of $T_c$ (vertical axis) $vs.$ lattice parameter $a$ (horizontal axis). Three distinct regions exist, representing three different materials UBe$_{13-x}$ (red, present study), UBe$_{13-x}$Al$_x$ (orange \cite{Amon2018b}), and U$_{1-y}$Th$_y$Be$_{13}$ (yellow \cite{Scheidt1998a, Shimizu2025}) -- all showing superconductivity, albeit of different character. In some cases, imperfections on the atomic scale can only be seen by means of transmission electron microscopy (insets).}
    \label{Summary}
\end{figure*}

Typically, unconventional superconductors are thought to be exceptionally sensitive to crystallographic imperfections and non-magnetic disorder \cite{Prozorov2024, Korshunov2016, Alloul2024, Gui2021}. For many uranium-based systems, this is also the case -- minute changes on the atomic level make a big difference in their ground states -- this can likely be linked to the oxidation state of uranium which is very sensitive to its immediate crystallographic environment \cite{Svanidze2025}. Crystallographic defects are minute and diverse: stacking faults in UPt$_3$ \cite{Aronson1990, Kim1997, Kycia1998}, spacial inhomogeneities of URu$_2$Si$_2$ \cite{Butch2012, Matsuda2011, Matsuda2008, Mydosh2011, Gallagher2016}, as well as uranium deficiencies coupled with local breaking of translational symmetry in UTe$_2$ \cite{Haga2022, Rosa2022, Sakai2022, Svanidze2025, Thomas2021a, Weiland2022}. UBe$_{13}$ single crystals, grown from aluminum flux, were conveniently available just when the interest in this material first arose \cite{Ott1983, Maple1985}. However, it took considerable effort to realize and prove that these crystals are in fact never UBe$_{13}$, but rather UBe$_{13-x}$Al$_x$ \cite{Volz2018, Amon2018b}, i.e. that the aluminum from the flux growth enters the crystal structure at the beryllium position -- see the orange region of "unintentional substitution" of Fig.~\ref{Summary}. With subsequent temperature treatment, the aluminum atoms start to leave the structure and, with longer annealing, vacancies -- as a result of aluminum removal -- appear, producing UBe$_{13-x}$. The respective amounts are small -- on the order of 0.1 at.\%, but superconductivity of this material changes not only in the value of $T_c$ as well as the upper critical field curve \cite{Langhammer1998}, but also sometimes revealing two superconducting anomalies \cite{Amon2018b}, making UBe$_{13}$ somewhat similar to UTe$_2$ (although the imperfections in the latter are even more subtle \cite{Svanidze2025}). In UBe$_{13-x}$Al$_x$, the partial substitution of beryllium by aluminum is also accompanied by an increase in lattice volume ($\sim 0.05$\%, see Fig.~\ref{Summary}) -- this makes it somewhat easier to track microscopic imperfections quantitatively. Furthermore, the change in volume, induced by unintentional substitution of aluminum from flux, is of similar magnitude to the case when uranium is substituted by thorium \cite{Smith1992, Smith1985, Jin1994} -- see the yellow region in Fig.~\ref{Summary}. Remarkably, both in UBe$_{13-x}$Al$_x$ (orange region) and in U$_{1-y}$Th$_y$Be$_{13-x}$Al$_x$ (yellow region) show double superconducting transitions. So, in principle, this could be related to crystallographic strain or disorder. Gallium substitution on the beryllium site also leads to behavior similar to the aluminum substitution \cite{Giorgi1984}.

In this sense -- especially considering that currently there is no other established way to grow single crystals of UBe$_{13}$ -- the only way to prepare pure, i.e. flux-free UBe$_{13}$ is by crystallizing the binary melt by which one obtains polycrystalline products. This work combines careful synthesis of UBe$_{13}$ with micro-scale isolation of single crystallites, giving access to the nominally cleanest possible UBe$_{13}$ so far -- both in terms of structural imperfections and single-phase character. 

\begin{figure*}
    \centering
    \includegraphics[width=\linewidth]{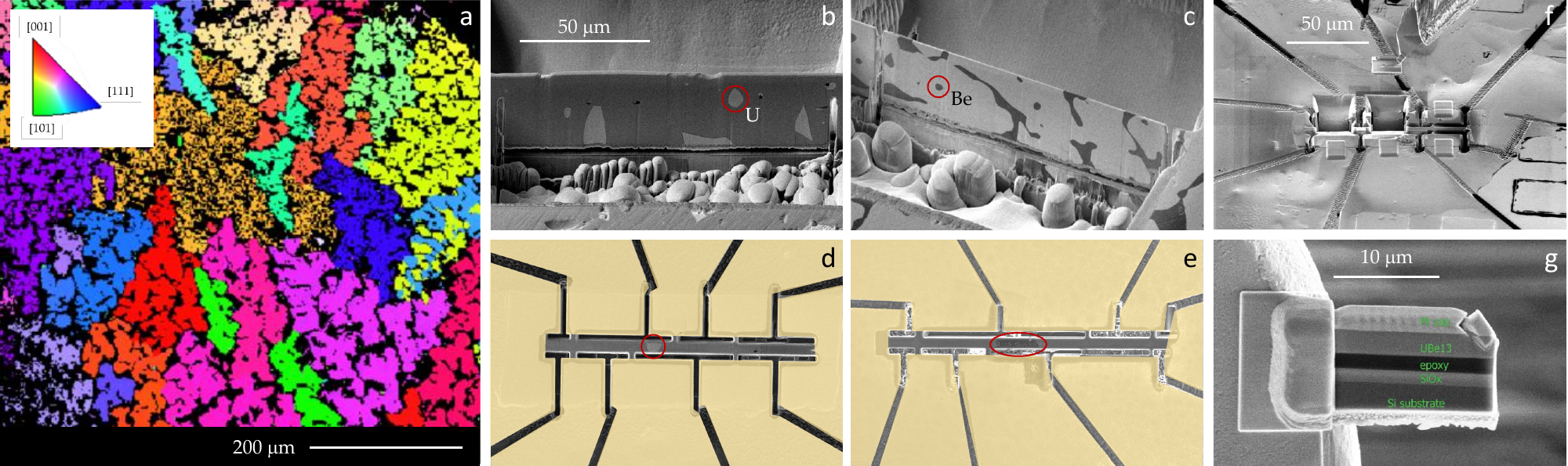}
    \caption{Extraction of intrinsic electrical resistivity of UBe$_{13-x}$: (a) Electron back-scatter diffraction is used to estimate grain size. (b) and (c): Inclusions of U (light gray) or Be (dark gray) secondary phases is easily recognizable in the lamellae, extracted from polycrystalline material. (d) and (e): Micro-scale devices, fabricated from polycrystalline samples. The current flow is left to right, which means that several voltage pairs avoid inclusions of U or Be -- this allows to capture the intrinsic resistivity of UBe$_{13-x}$. (f) and (g): A piece from the same device is then used to pinpoint defects on the atomic scale by means of transmission electron microscopy.}
    \label{Micro}
\end{figure*}

As demonstrated in this work, even in single-phase UBe$_{13}$ some beryllium vacancies can exist, as a result of the synthetic route. Hence, we now refer to those crystals as UBe$_{13-x}$. While from the chemical point of view the amount of this imperfection is non-trivial to quantify, the effect on the superconductivity is dramatic. For the lattice change on the order of 0.006\%, the corresponding change in $T_c$ is 28\%. In comparison, a similar change in $T_c$ for YBa$_2$Cu$_3$O$_{7-\delta}$ is only possible when the unit cell volume is changed by 0.5\% \cite{Schweiss1994}. Even in another uranium-based heavy-fermion superconductor UTe$_2$, the volume change of 0.09\% splits one transition into two, while $\sim 0.5$\% volume change kills superconductivity completely \cite{Svanidze2025}. Even when high-energy electron irradiation is employed to gently perturb the lattice of unconventional superconductors, stronger disorder is typically induced with small effects on $T_\mathrm{c}$ \cite{Alloul1991, Prozorov2014, Roppongi2023, Ghimire2024, Ruf2024, Ranna2025}. These examples show how large the effects of minute changes in lattice parameter on the critical temperature are in UBe$_{13}$, placing the latter in a separate paradigm -- even among unconventional superconductors. %This motivates continued search for this type of exotic phenomena in the lesser-explored space of uranium-based quantum materials.

\vspace{-10pt}
\section{Results and discussion}
\vspace{-10pt}

\begin{figure*}
    \centering
    \includegraphics[width=\linewidth]{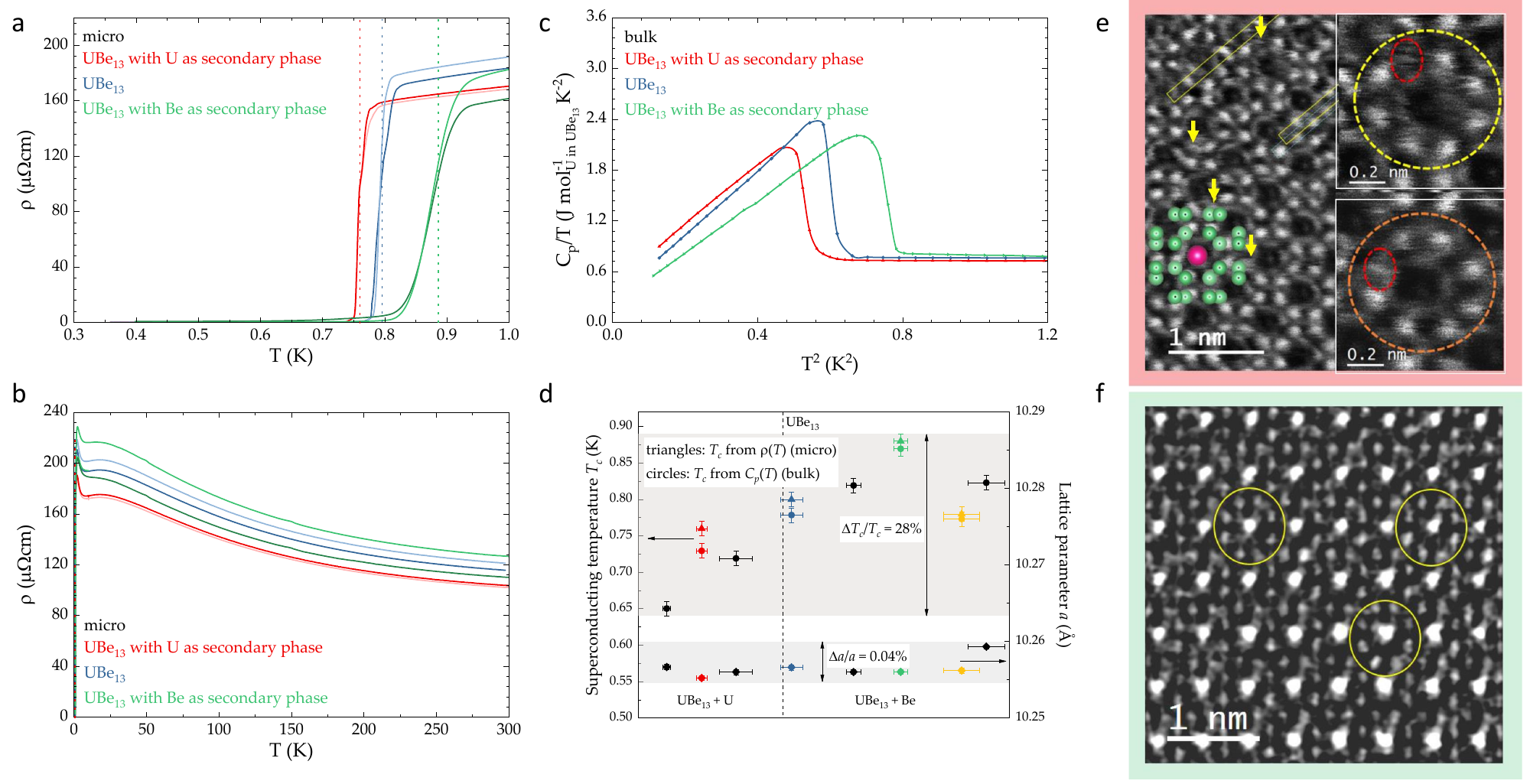}
    \caption{Intrinsic superconductivity of UBe$_{13-x}$: (a) and (b) Micro-scale resistivity reveals that the superconducting transition temperature of UBe$_{13-x}$ is changing by almost 30\%, depending on whether polycrystalline material is growing from U (red) or Be (green) excess. For all samples, only one superconducting anomaly is observed, with its position being different for different starting U:Be ratios. (c) The specific heat data follow the same trend. (d) All of the UBe$_{13-x}$ grains within a given sample have the same composition, given the agreement between the value of $T_c$ extracted from the micro-scale (triangles, resistivity) and bulk (circles, specific heat) data. Corresponding change in the lattice parameter $a$ is negligible (right axis). (e) and (f): The only difference is seen in the transmission electron microscopy -- revealing imperfect UBe$_{13-x}$ (red, top) and and perfect UBe$_{13}$ (green, bottom) material. (e) and (f) shows the sample with red and green markers in panel (d), respectively. The exact amount of this beryllium deficiency is estimated to be below 0.1\%.}
    \label{Tc}
\end{figure*}

\noindent $Structural~characterization$: Polycrystalline samples UBe$_{13-x}$ contain U or Be as a secondary phase, depending on the starting material composition. Physical properties of polycrystalline UBe$_{13-x}$ are affected by the grain boundaries, produced by both types of minority phases --  see Fig.~\ref{Micro}. The grains of UBe$_{13-x}$ are in fact fairly large, at least 50 $\mu$m in size, as evident from Fig.~\ref{Micro}(a). By using a focused ion beam patterning, we create single-grain devices from these samples. A careful device design \cite{Amon2019,Antonyshyn2020} can circumvent U or Be inclusions -- see Fig.~\ref{Micro}(b-e) -- giving access to the intrinsic resistivity of UBe$_{13-x}$. Contrary to previous studies, stoichiometries far from 1:13 have been considered in this work -- the nominal elemental ratios start with UBe$_{7.56}$ (UBe$_{13-x}$ with U as the secondary phase) and go to UBe$_{23.76}$ (UBe$_{13-x}$ with Be as the secondary phase). An unexpected result of this exploration was the large variation in the value of critical temperature $T_c$ for UBe$_{13-x}$ grown from either U or Be excess. By significantly changing the starting U to Be ratio for arc-melted UBe$_{13-x}$ samples, the value of $T_c$ is changed by nearly 30\% -- namely from 0.65 K to 0.88 K -- see Fig.~\ref{Tc}. The corresponding change in the lattice parameter $a$ for the whole range is 0.04\%, while the difference for the samples with highest (green) and lowest (red) $T_c$ is only 0.006\%. Overall, the unit cell parameter $a$ is staying virtually the same -- see the red region of Fig.~\ref{Summary}. This region is in stark contrast to the orange and yellow regions of the same figure. Here, the big difference in $T_c$ of UBe$_{13-x}$ cannot be explained by the presence of minority components such as aluminum, gallium or thorium. What is revealed by the transmission electron microscopy (Fig.~\ref{Tc}(e) and (f)) on pieces extracted from the micro-scale devices (Fig.~\ref{Micro}(f) and (g)) is that the amount of Be in the UBe$_{13-x}$ structure varies slightly. Quantification of Be by means of traditional chemical analysis is not always trivial \cite{Eckert2020, Buchner2020, Buchner2019,Bruce2011,Naglav2016} and in this study, single crystal analysis was not possible since a single-phase specimen of a suitable size ($20 \times 20 \times 20$ \AA$^3$) could not be isolated. Nonetheless, the transmission electron microscopy coupled with lattice parameter refinements based on powder data give an insight into the structural variations in UBe$_{13-x}$. It is important to note that while the importance of Be in UBe$_{13}$ was suspected, it has not been investigated much up until now \cite{Smith1985,Kim1992,Kim2022,Amon2018b}. It is of course important to understand why removal of less than 1 out of 24 Be atoms (average observed by TEM) within the environment of U in UBe$_{13-x}$ makes such a crucial difference -- again, quantitatively speaking it is hard to put a number on the abundance of this defect. Comparing with UTe$_2$ where the lattice parameter change on the order of 0.09\% arises from a defect with on the order of 0.01 \cite{Svanidze2025}, it is reasonable to suppose that the amount of beryllium deficiency in UBe$_{13-x}$ is even less. In principle, with the U-Be distances ($d_{U-Be} = 3.012$ \r{A}) being much smaller than the U-U distances ($d_{U-U} = 5.128$ \r{A}), the structure of UBe$_{13-x}$ can be represented by U-Be polyhedra with coordination of 24. In this regard, it is instructive to examine what happens in conventional superconductors with cage-like structures.

A strong sensitivity to small changes in the lattice has been observed in several cage-like superconductors, including YB$_6$ \cite{Sluchanko2017,Lavroff2024,Pribulova2023,Lortz2006,Schell1982}, Lu/YB$_{12}$ \cite{Sluchanko2011,Dudka2017,Czopnik2005}, Th$_4$Pt$_{16}$Be$_{33}$ \cite{Kozelj2021,Shang2023,Svanidze2021,Amon2020}, LaRu$_4$As$_{12}$ \cite{Mizukami2020}, Ba$_8$Ge$_{43}$ \cite{Carrillo2004, Shimizu2007}, Ba$_6$Ge$_{25}$ \cite{Yuan2002, Yuan2004, Carrillo2005}, and Ba$_8$Si$_{46}$ \cite{Castillo2015,Tanigaki2003,Yamanaka2010, Liang2011,Lory2017}. Here, small changes in the occupancy of the cage drive changes in the atomic environment at the center of the cage, drastically influencing the bonding within the structure. In principle, for phonon-mediated superconductivity, which is driven by (i) density of states at the Fermi level and (ii) electron-phonon coupling, structural change will directly affect phonon frequencies. In fact, it has been shown previously that "rattling" can enhance both superconductivity \cite{Winiarski2016,Yamaura2006,Zenji2005,Hiroi2011,Fijalkowski2021} as well as effective electron mass \cite{Sanada2005, Bauer2002} -- in general, the larger the cage, the stronger the "rattling" \cite{Matsuhira2009, Lory2017}. In this sense, both superconductivity and enhanced effective electron mass of UBe$_{13-x}$ could, in principle, originate from "rattling". However, crystallographic analysis of UBe$_{13-x}$ \cite{Goldman1985}, as well as other related compounds with the NaZn$_{13}$-type cubic structure \cite{Hidaka2018} did not point to rattling, consistent with superconductivity of UBe$_{13-x}$ not being phonon-driven. Another parameter is likely at play -- namely the oxidation state of uranium. This has also been stipulated to be the reason for the $T_c$ suppression in UTe$_2$ \cite{Deng2024, Svanidze2025}. This scenario is supported by the magnetic susceptibility data, shown in Fig.~S2\cite{SI} -- both effective magnetic moment and Weiss temperature, extracted from the Curie-Weiss fits, change for different UBe$_{13-x}$ samples, showing that the local environment has a big effect on the $f$-electron state. It is important to note that the previously reported anomaly in magnetic susceptibility data around $T = 160$ K (single crystals of UBe$_{13-x}$Al$_x$ \cite{Thomas1998}) can likely arise due to the presence of Be-doped U (see Fig.~S2\cite{SI}). Nonetheless, the intrinsic magnetism of UBe$_{13-x}$ does appear to change -- the Curie-Weiss fitting in the 200-300 K temperature range reveals that the effective magnetic moment ($\mu_{eff}$) and magnetic correlation strength (as reflected by the Weiss temperature $\Theta_W$) are actually the smallest for the sample with full Be occupancy and highest $T_c$ (green symbols). It is also likely that the valence electron concentration in the Be framework of UBe$_{13-x}$ is able to accommodate these valence fluctuations of U -- unlike in UB$_{12}$ \cite{Troc2015}, where no superconductivity is observed even though structurally these two materials are quite similar. Another possible effect of Be variation in UBe$_{13-x}$ is a very small (beyond detectable) change in the cage volume, i.e. application of minuscule pressure. This can, in principle, have rather big effects on the band structure and effective masses that could affect the superconducting state. Though we argue below that superconductivity of UBe$_{13-x}$ does not appear to be extremely sensitive to pressure tuning.

\noindent $Strain~effects$: Since UBe$_{13-x}$ crystallites are always surrounded by a secondary phase of either Be or U, which contracts upon cooling differently from UBe$_{13-x}$, it is necessary to examine the possible effect of strain. It is known that $T_c$ of unconventional superconductors can be very sensitive to strains, e.g. in ruthenate \cite{Hicks2014} or iron-based superconductors \cite{Malinowski2020}. For Co-doped CaFe$_2$As$_2$, it was demonstrated that the superconducting properties can even be effectively manipulated through the strain imposed by flux inclusions \cite{Ran12,Gati12}. Using previous reports of the uniaxial pressure sensitivity, derived from thermal expansion, d$T_c$/d$p_{a} \approx 4 $ mK/kbar and a bulk modulus of 108 GPa \cite{Benedict1987}, we estimate the sensitivity of $T_c$ to strain, $\varepsilon_a:=(\Delta a/a)$ to be d$T_c$/d$\varepsilon_{a} \approx 4$ K. Thus, in order to explain a shift of $\approx$ 100 mK between single phase samples and samples with Be minority phase, a tensile strain of about 2.5 \% would be needed. However, the differential thermal expansion between UBe$_{13}$ and elemental Be (see SI) is vanishingly small and results in tensile strains of the order of 0.01 \% only. Even the differential thermal expansion between UBe$_{13}$ and elemental U induces a maximum strain of about -0.15 \% at low temperatures. Moreover, micro-scale (possible strain due to substrate \cite{Bachmann2019}) and bulk (possible strain due to U/Be inclusions) trend in the evolution of $T_c$ are identical (triangles $vs.$ circles in Fig.~\ref{Tc}(d)). We therefore conclude that strain effects created by the U/Be inclusions cannot fully account for the shift of $T_c$, observed in this study.

\begin{figure}
    \centering
    \includegraphics[width=\linewidth]{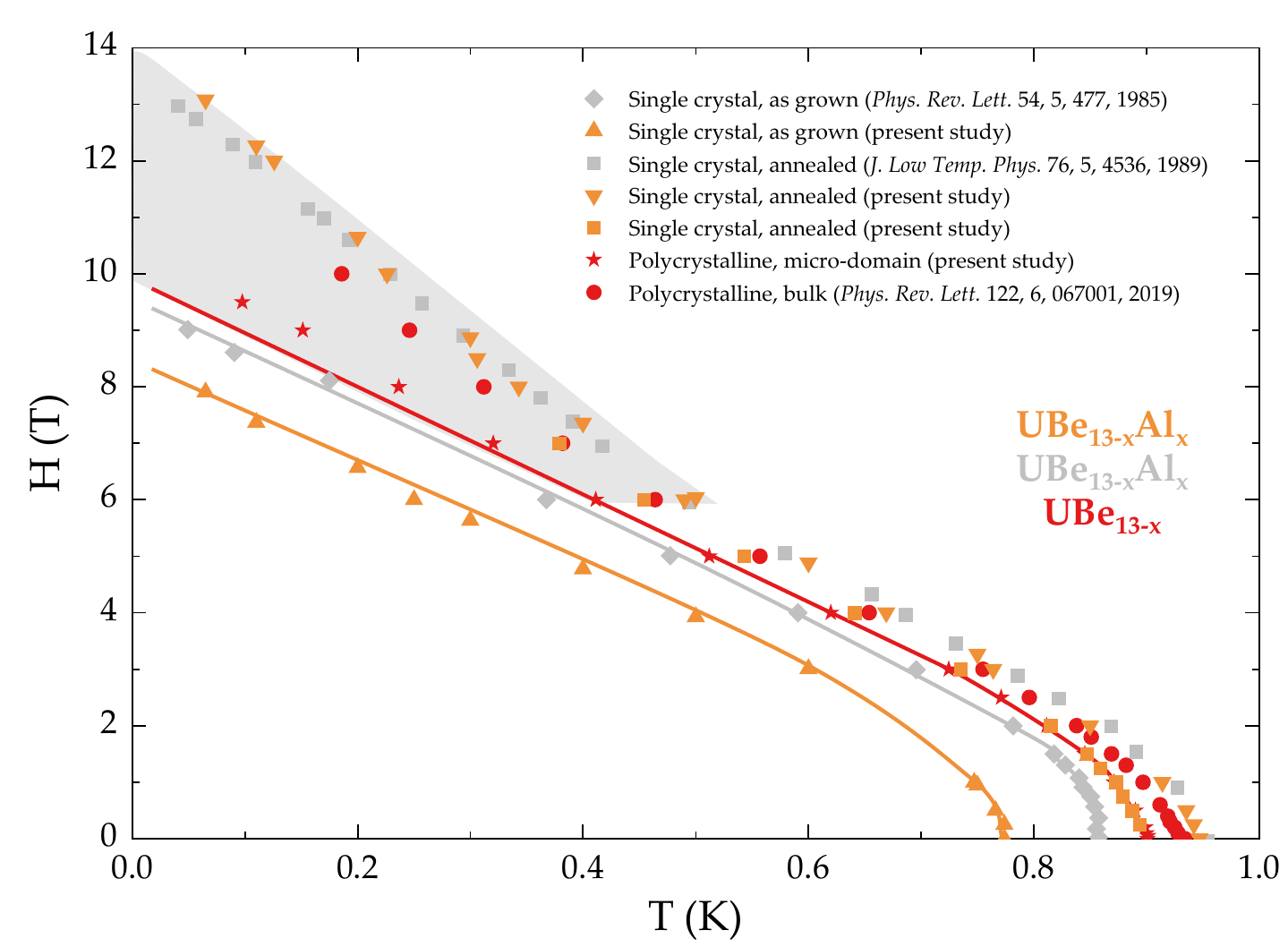}
    \caption{Comparison of critical fields of UBe$_{13-x}$Al$_x$ and UBe$_{13-x}$ from previous studies on single crystals (gray symbols \cite{Maple1985, Brison1989a}), together with the current work on single crystals (orange), polycrystalline micro-scale (red stars), and bulk (red circles) samples.}
    \label{Hc2}
\end{figure}

\noindent $Thermodynamic~measurements$: Overall, it appears that superconductivity in UBe$_{13-x}$ is very sensitive to the exact ratio of U and Be with a large variation of $T_c$. It is important to distinguish UBe$_{13-x}$ $vs.$ UBe$_{13-x}$Al$_x$ $vs.$ U$_{1-y}$Th$_y$Be$_{13}$ -- as summarized in Fig.~\ref{Summary}. In previous studies, a number of experimental results on aluminum-flux grown single crystals and arc-melted polycrystalline samples have been reported, and their conflicts regarding superconducting symmetries make it difficult to reach a coherent understanding: while specific heat analysis \cite{Jin1994, Ott1984}, Josephson tunneling \cite{Han1986}, and penetration depth measurements \cite{Einzel1986} have been performed on polycrystalline material (i.e. UBe$_{13-x}$), a number of studies on single crystals (i.e. UBe$_{13-x}$Al$_x$) \cite{Shimizu2012, Matsuno2015,Mayer1986, Hiess2014} should be considered separately. This raises the question of what kind of superconducting symmetry is realized in UBe$_{13}$ and if it is different from the one in UBe$_{13-x}$Al$_x$.

%From specific-heat measurements in rotating fields $C(T, H)$ using Al-flux grown single crystalline UBe$_{13}$, fully-gapped superconducting quasiparticle excitations have been observed \cite{Shimizu2015}. The present study is based on polycrystalline samples, for which such an analysis is not possible, making it difficult to conclude the presence or orientation of nodes in UBe$_{13}$. Nevertheless, the field dependence of the specific heat in arc-melted UBe$_{13}$ polycrystals provides valuable information \cite{Walti2001}. Although the authors argue for point nodes with a scaling analysis, the low-$H$ data show a linear $C(H) \propto H$, consistent with single-crystal results \cite{Shimizu2015}. This agreement supports a fully gapped superconducting state in  polycrystalline UBe$_{13}$ as well.

The upper critical field of UBe$_{13}$ is anomalous (see Fig.~\ref{Hc2}), far exceeding the BCS Pauli limit ($\mu_{0} H_{\rm P} \simeq$ 1.6 T), which supports an occurrence of a spin-triplet superconductivity \cite{Ott1984, Fomin2000, Shimizu2019, Minami2026}.  In the $O_{h}$ cubic symmetry fully gapped spin‑triplet representations exist \cite{Blount1985}: the fully gapped
$A_{1u}$ state or a mixed state of$A_{1u}$ and the (nematic) $E_{u}$
state, consistent with the nodeless  behaviour in low‑temperature specific‑heat \cite{Shimizu2015}. The pronounced curvature of $H_{c2}(T)$ near $T_{\mathrm c}$ indicates strong paramagnetic effects and an enhanced $H_{\rm P}$, which is also compatible with the $A_{1u}$ state \cite{Shimizu2011,Shimizu2016}. Furthermore, samples with a high $T_c$ exhibit an upturn of $H_{c2}$ at roughly $T/T_{c}= 1/2$ whereas lower-$T_c$ samples  do not \cite{Langhammer1998}. The origin of this upturn has been controversial \cite{Langhammer1998, Thomas1996, Schmiedeshoff1988, Brison1989a, Signore1995} and can in principle have various origins such as strong-coupling or multiband effects, or field-enhanced pairing interactions. A successful description combines a field-induced weak $E_u$ component mixed with the dominant $A_{1u}$ state. Our measurements of $H_{c2}$ on two single crystals, grown from aluminum flux, as well as those obtained on micro-scale domains of arc-melted polycrystals, shown in Fig. \ref{Hc2}, confirm this trend without a noticeable difference between UBe$_{13-x}$ and UBe$_{13-x}$Al$_x$. Although the dependence of the superconducting transition temperature $T_c$ on the lattice parameter $a$ differs systematically between UBe$_{13-x}$ and UBe$_{13-x}$Al$_x$ (Fig.~\ref{Summary}), the impacts of beryllium vacancies versus aluminum inclusions on the critical field curve in UBe$_{13}$ seems to be similar.

%Recent pressure experiments on arc-melted UBe$_{13}$ \cite{Shimizu2019} also give similar results as pressure studies on Al-flux grown single crystals \cite{Fomin2000}: While for pressures below 3 GPa  the spin-triplet $A_{1u}$ state is dominant, causing the strong bending of $H_{c2}$, and the weaker $E_{u}$ state is mainly induced by magnetic field causing the anomalous enhancement of $H_{c2}$ at $T/T_{c}= 1/2$, the latter is enhanced for higher pressure in a way that the strong bending disappears. In particular, it is noteworthy that the strong curvature of $H_{c2}$ at low fields near $T_\mathrm{c}$ cannot be reproduced at ambient pressure without considering the $A_{1u}$ state ($\bm{\Psi} = \bm{\hat{x}} k_x + \bm{\hat{y}} k_y + \bm{\hat{z}} k_z$). The $E_{u}$ representation is two-dimensional, and a fully gapped state ($\bm{\Psi_{1}} = \bm{\hat{x}} k_x + \bm{\hat{y}} k_y - 2 \bm{\hat{z}} k_z$) and a point-nodal state ($\bm{\Psi_{2}} = \bm{\hat{y}} k_y - \bm{\hat{x}} k_x$) are degenerate \cite{Blount1985, Fomin2000}. In the high-pressure region where $E_{u}$ is favored, the point-nodal $E_{u}$ ($\bm{\Psi_{2}}$) state may be masked by other fully gapped superconducting states, if present.

\vspace{-10pt}
\section{Summary and outlook}
\vspace{-10pt} 

It is clear that UBe$_{13}$ is an unconventional superconductor, which makes its ground state highly susceptible to atomic-scale imperfections on both U and Be sites. This material is also one of the oldest spin-triplet candidates, which makes retrospective analysis of lattice quality $vs.$ observed phenomena much needed. In this letter we establish an extreme sensitivity of UBe$_{13}$ to beryllium defects -- the removal of less than 0.1\% of beryllium does not change unit cell volume, but reduces $T_c$ by nearly 30\%. It is important to note that in the UBe$_{13}$ sample with the highest $T_c$, we do not find any defects (beryllium or other kind), despite implementing the same meticulous analysis on both macro and atomic scales. The numerous TEM images, obtained as part of this study, do not show any indications of imperfections. While it is inherently difficult to prove crystallographic perfection, the samples of UBe$_{13}$ with the highest $T_c$ are different from both polycrystalline and single crystalline samples of UBe$_{13}$ for which we conclusively show defects. Even for an unconventional superconductor, such a delicate structure-properties relation is unprecedented. 

The similar trend between $T_\mathrm{c}$ and the upper critical field curve of UBe$_{13-x}$ and UBe$_{13-x}$Al$_x$ might be explained with a similarity of the samples of highest $T_\mathrm{c}$. Since annealing of Al-grown single crystals removes Al from the structure and leaves Be vacancies \cite{Amon2018b}, it seems that high-$T_\mathrm{c}$ samples from both growth methods share a small amount of Be vacancies. In the picture of admixed $A_{1u}$ and $E_{u}$ triplet superconducting states in UBe$_{13}$\cite{Fomin2000, Shimizu2019}, variations in their relative weight explain the observed trends: Since the $A_{1u}$ state is a fully gapped superconducting state with a well-defined phase, it is sensitive to imperfections, but likely more robust than the lower symmetry order parameter such as $E_{u}$. Thus, for higher $T_c$ samples, $A_{1u}$ prevails, and a weaker $E_{u}$ leads to the  anomalous upward curvature in $H_{c2}$ for $T_{c}/2$. For samples with lower $T_c$, imperfections lead to a suppression of both components with a stronger reduction of the $E_u$ state, so that the upturn disappears. The critical field curve of the micro-scale UBe$_{13}$ (stars in Fig.~\ref{Hc2}) is in between those extreme cases and shows a small upturn. Hence, the imperfections leading to this relation between the upturn and $T_c$ seem to be similar in aluminum-free polycrystals with beryllium vacancies and single crystals with aluminum inclusions.
%Overall, it appears that the intrinsic UBe$_{13}$ -- that is the binary compound in which there are no observable beryllium vacancies -- still has an upturn in the $H-T$ curve. This is supporting the hypothesis that $A_{1u}$ and $E_{u}$ triplet superconducting states coexist in UBe$_{13}$, indicating that it is likely an intrinsic spin-triplet superconductor in its unperturbed state.

\noindent $Acknowledgments$: E.S. thanks the Christiane N{\"u}sslein-Volhard-Stiftung. E.S., M.K., N.Z. and P.K. acknowledge the support of the Boehringer Ingelheim Plus 3 Program. N.Z. is grateful for the support of the Alexander von Humboldt Foundation through the Philipp Schwartz Initiative. E.G. gratefully acknowledges the funding through the Deutsche Forschungsgemeinschaft (DFG, German Research Foundation) Grant No. TRR 288-422213477 (Project No A13). Additionally, E.H. acknowledges funding by the DFG through CRC1143, Grant No 247310070 (Project No. C10) and the W\"urzburg-Dresden Cluster of Excellence on Complexity and Topology in Quantum Matter—ct.qmat (EXC 2147, Project ID 390858490).

\noindent $Data ~ availability$: The data that support the findings of this article are available from the authors upon reasonable request.

%An open question is how Al-flux grown samples with visible Al inclusions (\ref{Tc}) can have critical temperatures as high as those in Al-free polycrystals. Micro-scale isolation of small single-crystallites from polycrystalline material is used to  access UBe$_{13}$ material that is as close to the "perfect" structure as possible.

%Since $A_{1u}$ possesses antiparallel spin components, Pauli limiting in the case of $A_{1u}$ is more effective than in the $E_u$ case, leading to a reduced $H_{c2}$—consistent with experimental observations (see Fig.~\ref{Hc2}).

%As summarized in Fig.~\ref{Summary}, the smaller lattice parameter $a$ of UBe$_{13 - x}$, compared to UBe$_{13 - x}$Al$_{x}$ suggests that UBe$_{13 - x}$ is closer to the high-pressure regime, in which $E_{u}$ contributions become more prominent than that in UBe$_{13 - x}$Al$_{x}$. To further investigate the nature of the superconductivity in UBe$_{13}$, it is essential to perform pressure experiments \cite{Helm2020,Grockowiak2022,Sakai2015} on micro-scale specimens of UBe$_{13}$, since bulk samples remain out of reach.

% In fact, several criteria \cite{Stewart2019} that argue for unconventional superconductivity of UBe$_{13-x}$ should be re-visited: while specific heat analysis \cite{Jin1994, Ott1984}, Josephson tunneling \cite{Han1986}, and penetration depth measurements \cite{Einzel1986} have been performed on polycrystalline material, i.e. UBe$_{13-x}$, a number of studies on single crystals (i.e. UBe$_{13-x}$Al$_x$) \cite{Shimizu2012, Matsuno2015,Mayer1986, Hiess2014} should be considered separately.

%\vspace{10pt} 
\bibliography{lit.bib}
\vspace{5pt} 

%%%%%%%%%%%%%%%%%%%%%%%%%%%%%%%%%%%%%%%%%%%%%%%%%%%%%%%%%%%%%%%%%%%%%%%%%%%%%%%%%%%%%
\section{Supplementary Information}

\renewcommand{\thetable}{S\arabic{table}}
\renewcommand{\thefigure}{S\arabic{figure}}
\setcounter{page}{1}
\setcounter{figure}{0}
\setcounter{table}{0}

\begin{center}
\textbf{Sample synthesis}
\end{center}

\begin{figure}[htbp]
    \centering
    \includegraphics[width=\linewidth]{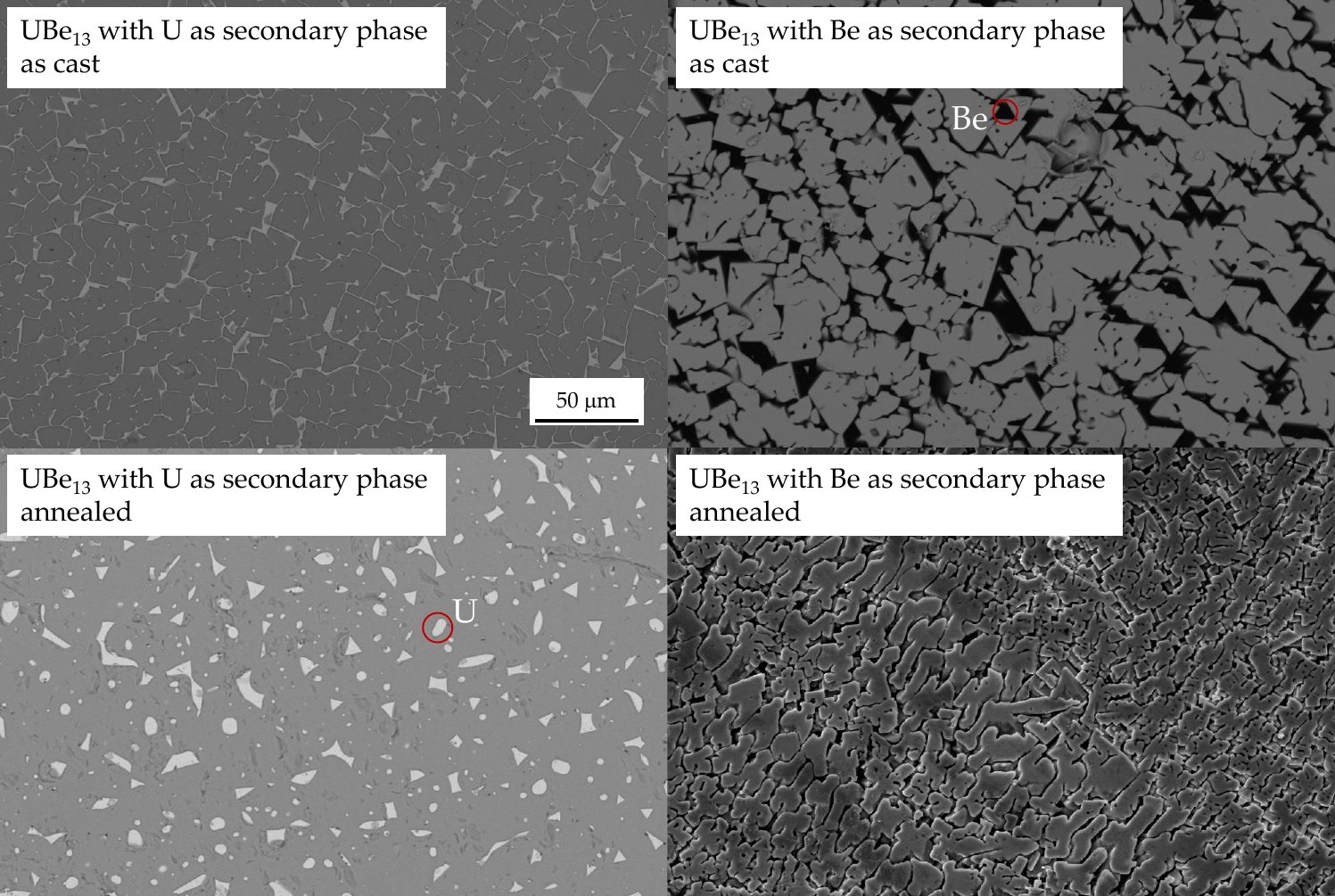}
    \caption{Microstructure of UBe$_{13-x}$: In order to access UBe$_{13-x}$ in its purest form, polycrystalline material with varying starting U:Be ratio was prepared. In as-cast material (top panels), right angles are reflecting crystallites growing in cubic symmetry (the NaZn$_{13}$ structure type, $Fm\bar{3}$c space group). Grain domain boundaries are either U (white) or Be (black). With annealing (bottom panels), grains of UBe$_{13-x}$ grow and grain boundaries diminish.}
    \label{EDX}
\end{figure}

All sample preparation and handling was performed in the specialized laboratory, equipped with an argon-filled glove box system, dedicated to the handling of Be-containing samples (MBraun, p(H$_2$O/O$_2$)$<$ 0.1 ppm) \cite{Leithe-Jasper2006}. Polycrystalline sample of UBe$_{13-x}$ with varying U:Be ratio were synthesized by arc melting from elements Be (Heraeus, $>99.9$ wt.\%) and U (sheets, Goodfellow, 99.98\%). Total of eight samples, shown in Fig.~3, spanned the starting U:Be ratio from UBe$_{7.56}$ to UBe$_{23.76}$. Due to high vapor pressure of Be, some of it is always lost during arc-melting. The mass difference between the pre- and post-arcmelted samples is used to estimate horizontal error bars in Fig.~3. The arc-melted samples were subsequently annealed, by placing them inside a BeO crucible, inside a sealed Ta tube in an inert Ar atmosphere. The U-rich samples were annealed at 1100$^\circ$C, while Be-rich samples were annealed at 1250$^\circ$C for one week. A piece of Be was added to avoid Be evaporation, however, in Be-rich samples, some Be is still lost from the sample surface (see bottom right panel in Fig.~\ref{EDX}). This loss, however, did not result in a noticeable mass difference. For both U- and Be-rich samples, annealing promoted significant grain growth, as evident from the micro-structure analysis -- see Fig.~\ref{EDX}.

\begin{center}
\textbf{Chemical analysis}
\end{center}

\begin{figure*}
    \centering
    \includegraphics[width=\linewidth]{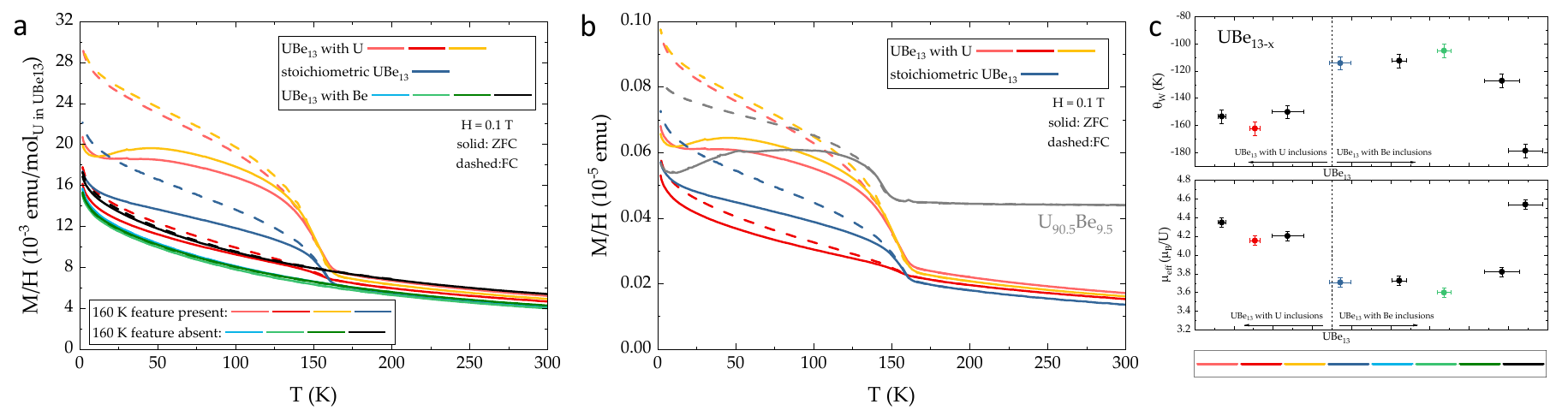}
    \caption{Magnetic properties of bulk UBe$_{13-x}$: (a) Temperature-dependent susceptibility data indicate that both effective magnetic moment and Weiss temperature change for samples with different Be occupancy. (b) The transition around $T = 160$ K is likely coming from Be-doped U secondary phase (gray line), which does not follow a Curie-Weiss law. (c) The values of the Weiss temperature and effective paramagnetic moment, extracted from the Curie-Weiss fits. Note that the maximum critical temperature $T_c$ is seen in the sample with the smallest moment and absolute value of $\Theta_W$ (green line and symbols).}
    \label{MT}
\end{figure*}

Powder X-ray diffraction was performed on a Huber G670 Image plate Guinier camera with a Ge-monochromator (CuK$_{\alpha_1}$ radiation, $\lambda$ = 1.54056 \r{A}) by mixing UBe$_{13-x}$ with a LaB$_6$ standard. Phase identification was done using the WinXPow software \cite{WinXPow}. The lattice parameters were determined by a least-squares refinement using the peak positions, extracted by profile fitting (WinCSD software \cite{Akselrud2014}). UBe$_{13-x}$ samples were additionally analyzed by energy-dispersive x-ray spectroscopy with a Jeol JSM 6610 scanning electron microscope equipped with an UltraDry EDS detector (ThermoFisher NSS7). The semi-quantitative analysis was performed with 30 keV acceleration voltage. No impurity elements were observed, confirming that no reaction with the crucible took place during synthesis. 

Spherical-aberration corrected high-resolution TEM (HRTEM) and scanning TEM (HRSTEM) analyses of the sample were performed with JEM-ARM300F microscope (Grand ARM, JEOL, Akishima, Japan) with double correction. Dodecapole correctors in the beam and the image forming system correct the spherical aberration of the condenser and the objective lenses, respectively. TEM resolution is 0.5-0.7 \r{A}, depending on resolution criterion applied, STEM resolution is 0.5 \r{A}. TEM images were recorded on a $4k \times 4k$ pixel CCD array (Gatan US4000).

%\begin{figure*}
%    \centering
%    \includegraphics[width=\linewidth]{Fig4.pdf}
%    \caption{Superconducting and normal state properties of bulk (top panels, specific heat) and micro-scale (bottom panels, resistivity) UBe$_{13-x}$. For both normal and superconducting states, a clear difference is seen.}
%    \label{Fig4}
%\end{figure*}

\begin{center}
\textbf{Bulk specific heat and magnetization analysis}
\end{center}

\begin{figure}
\includegraphics[width=8cm]{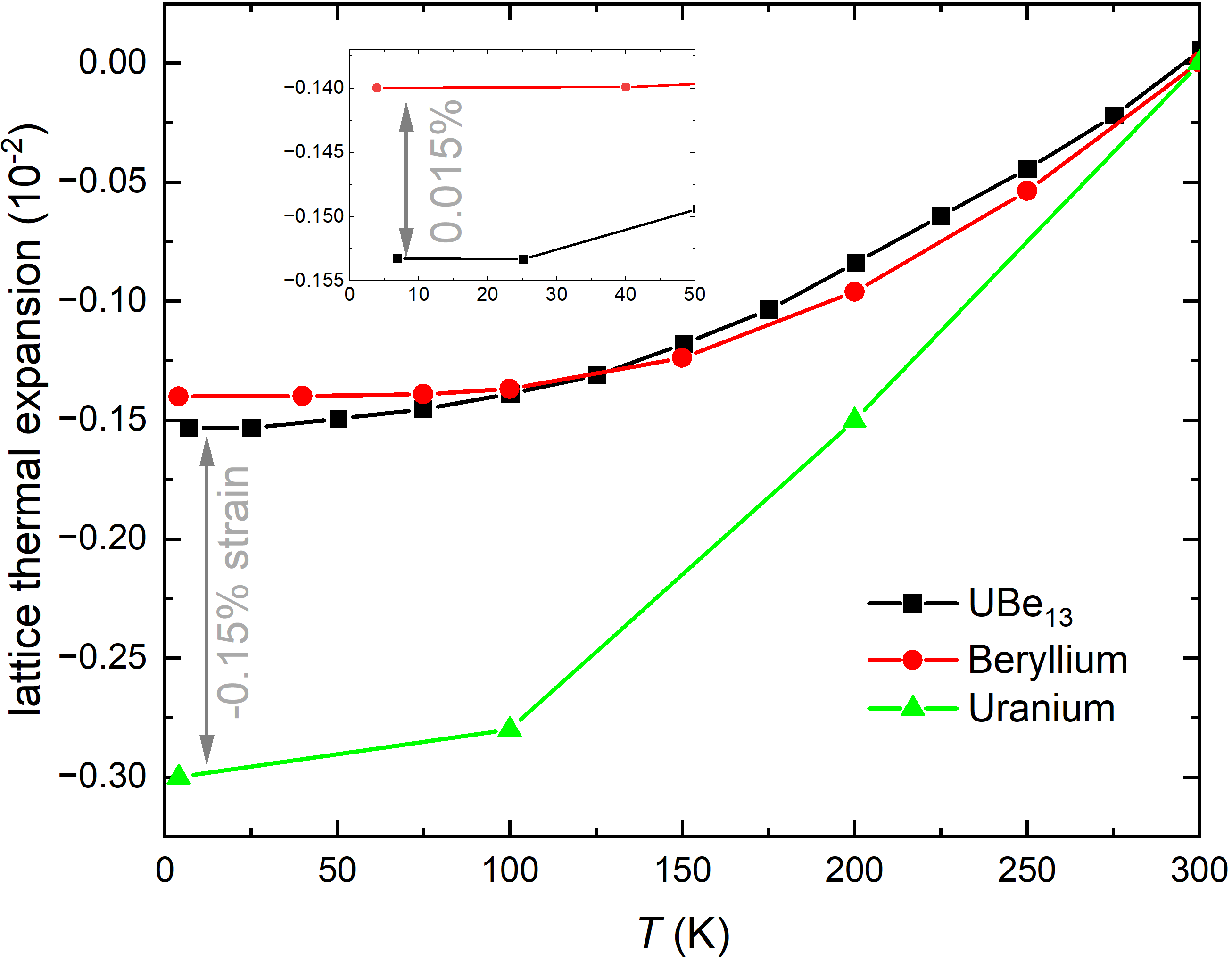}
\caption{Lattice thermal expansion as a function of temperature for UBe$_{13}$ (from Ref. \cite{Hidaka2018}), elemental Be (Ref. \cite{Bodryakov2018}) and elemental U (Ref. \cite{Schuch1952}). The arrows indicate the amount of strain that is induced by differential thermal expansion between UBe$_{13}$ and elemental U (main figure) or between UBe$_{13}$ and elemental U (inset).}
\label{fig:differential thermal expansion}
\end{figure}

Magnetic susceptibility and heat-capacity measurements on bulk UBe$_{13-x}$ samples were carried out on a Quantum Design Magnetic Property Measurement System (MPMS) and a Physical Property Measurement System (PPMS), respectively. For both, contributions coming from elemental U or elemental Be were subtracted from the as-measured data. The magnetic susceptibility data, summarized in Fig.~\ref{MT}, show the $T= 160$ K transition, reported previously \cite{Thomas1998}. This feature is only present for samples grown from U-excess -- see panel (b). It is likely that this feature is not intrinsic to UBe$_{13-x}$, but rather a result of some elemental U being doped by Be -- see gray curve in panel (b). However, it is important to note that the Curie-Weiss behavior, observed for all UBe$_{13-x}$ samples is likely intrinsic to the UBe$_{13-x}$ phase. As summarized in panel (c), the values of the effective moment $\mu_{eff}$ and Weiss temperature $\Theta_W$, extracted from the fits above 200 K evolve non-monotonously. Note that both the absolute value of $\Theta_W$ and the value of $\mu_{eff}$ are smallest for the sample with the highest $T_c$ value (green).

\begin{center}
\textbf{Resistivity measurements}
\end{center}

\begin{figure}
\includegraphics[width=8cm]{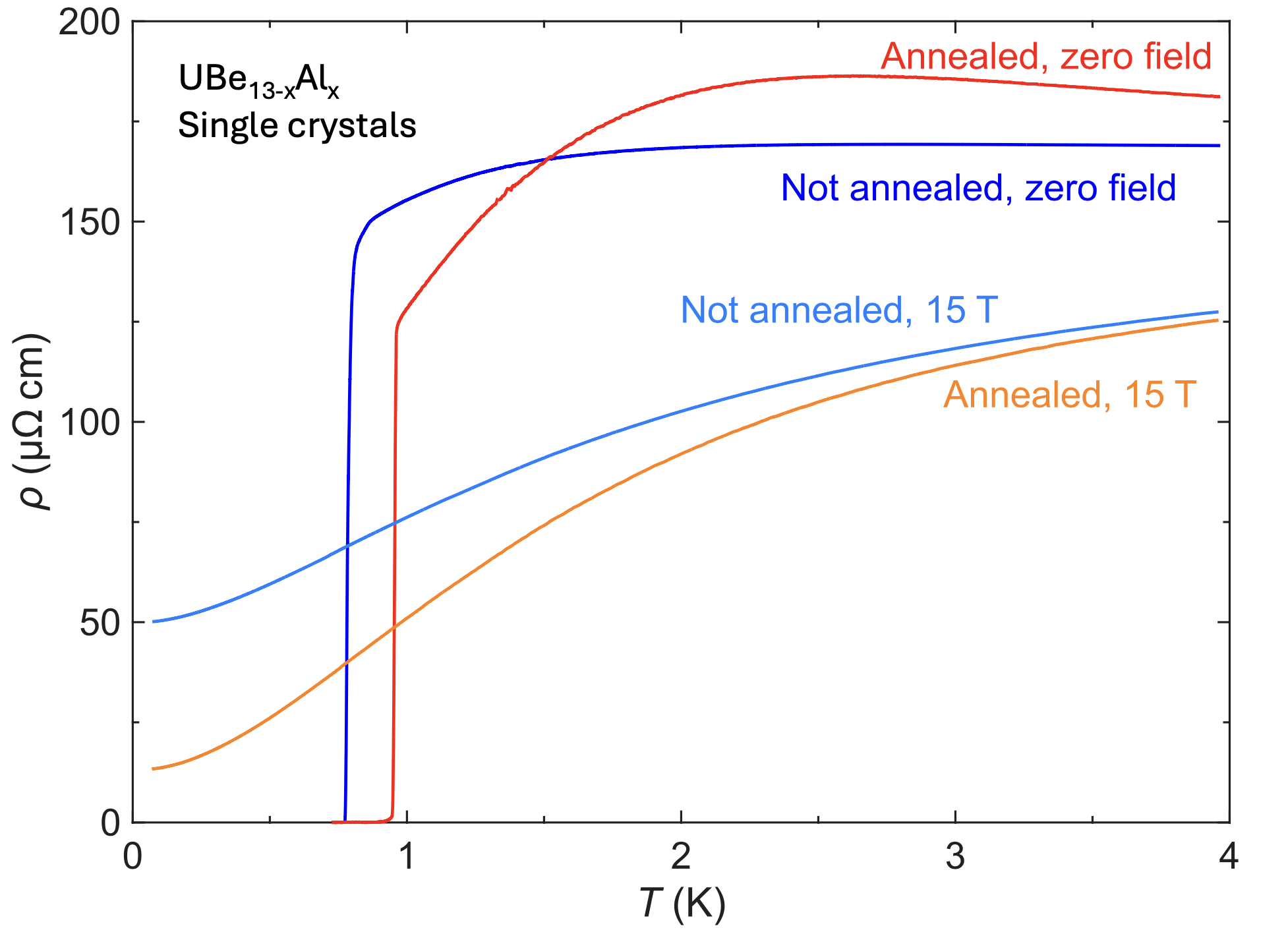}
\caption{Low-temperature resistivity for zero field and $\mu_0H = 15$ T of two single crystals of UBe$_{13-x}$Al$_x$, grown in aluminum flux (same samples as those used in Ref. \cite{Amon2018b}).}
\label{fig:rhovsTsc}
\end{figure}

\begin{figure}
\includegraphics[width=8cm]{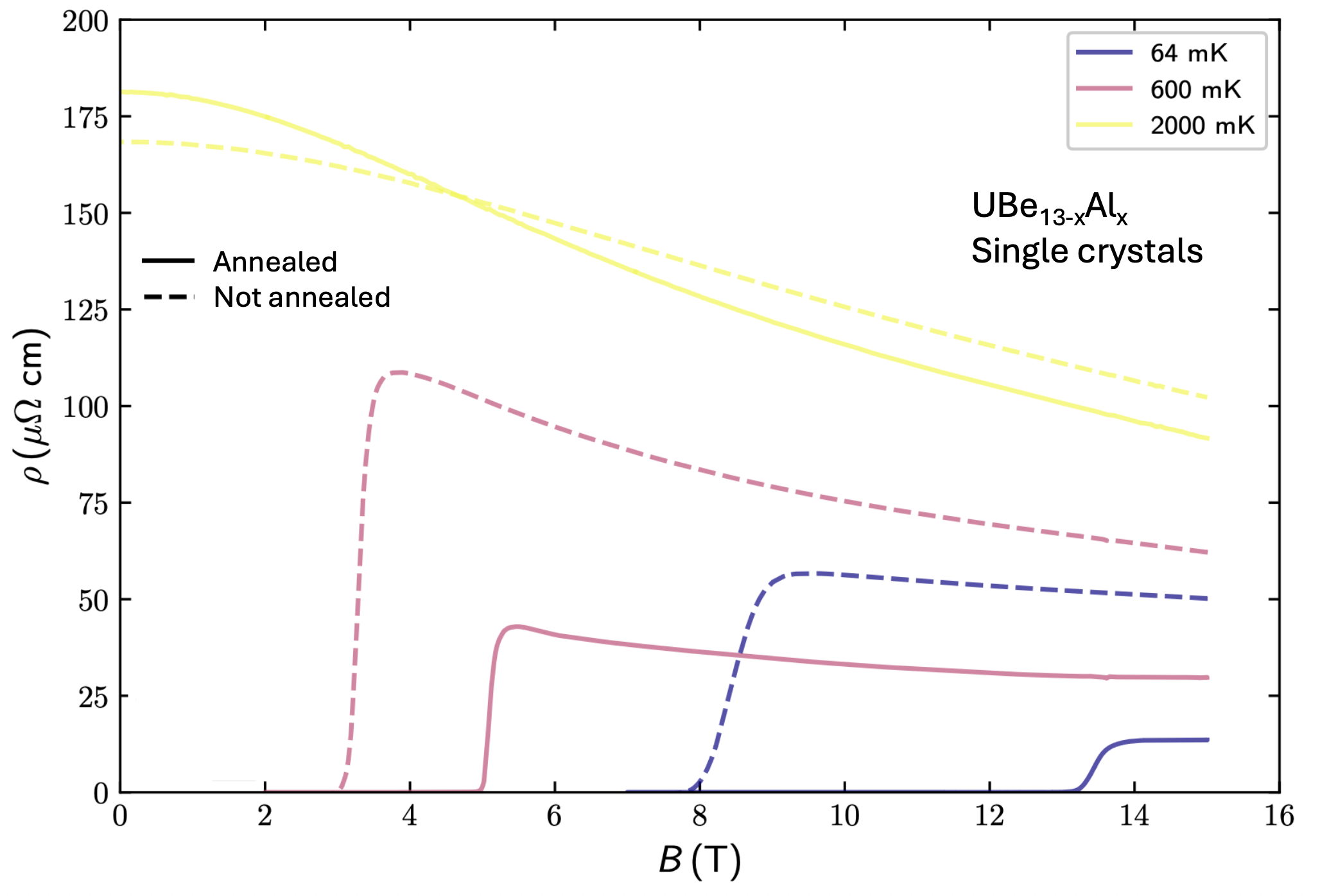}
\caption{Resistivity as a function of magnetic field at low temperature for UBe$_{13-x}$Al$_x$ -- the same samples as those shown in Fig.~\ref{fig:rhovsTsc}.}
\label{fig:rhovsHsc}
\end{figure}

\begin{figure}
\includegraphics[width=8cm]{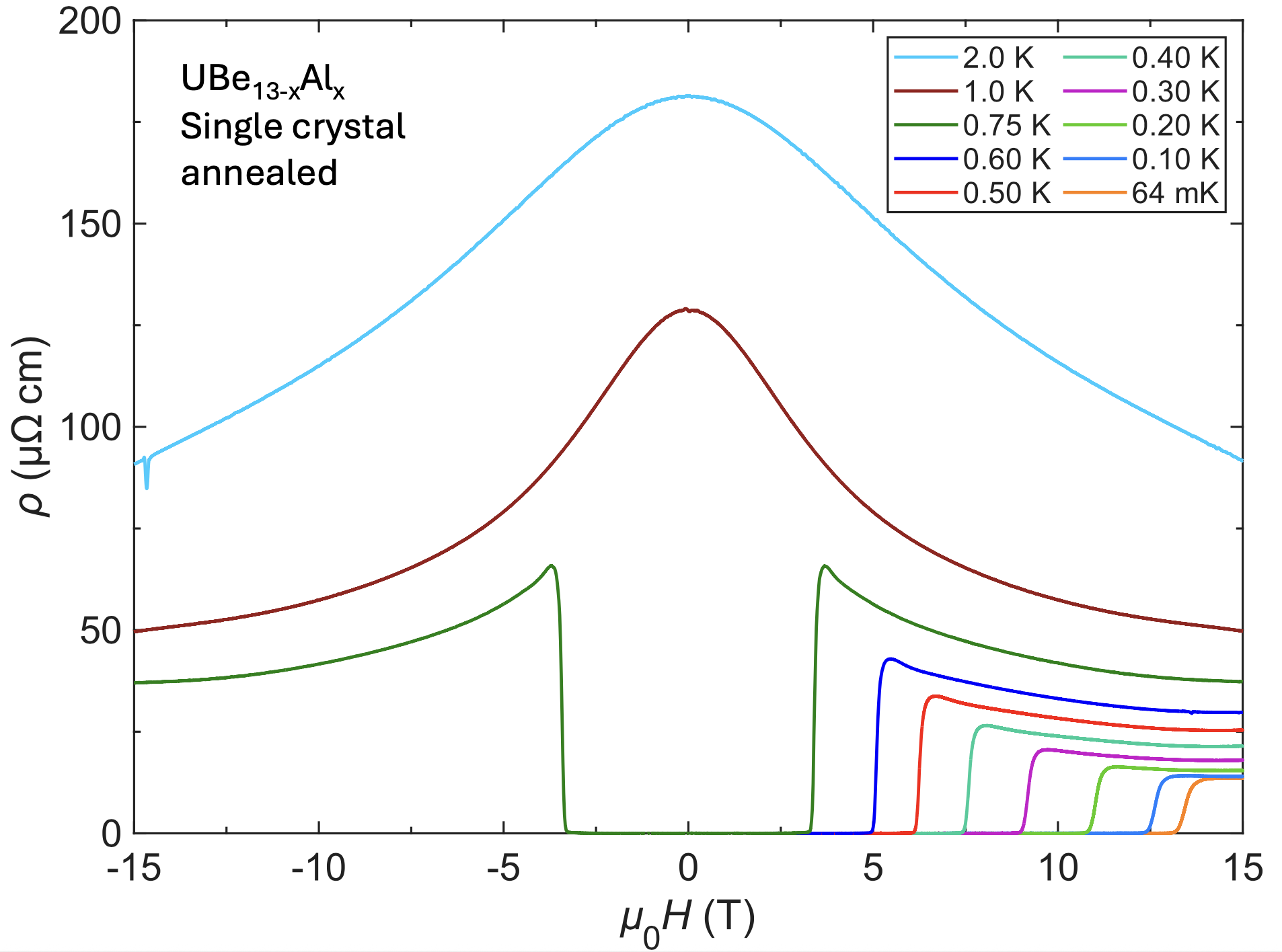}
\caption{More extensive dataset of resistance as a function of magnetic field at low temperature for the annealed UBe$_{13-x}$Al$_x$ sample.}
\label{fig:rhovsTmoredata}
\end{figure}

\begin{figure}
\includegraphics[width=8cm]{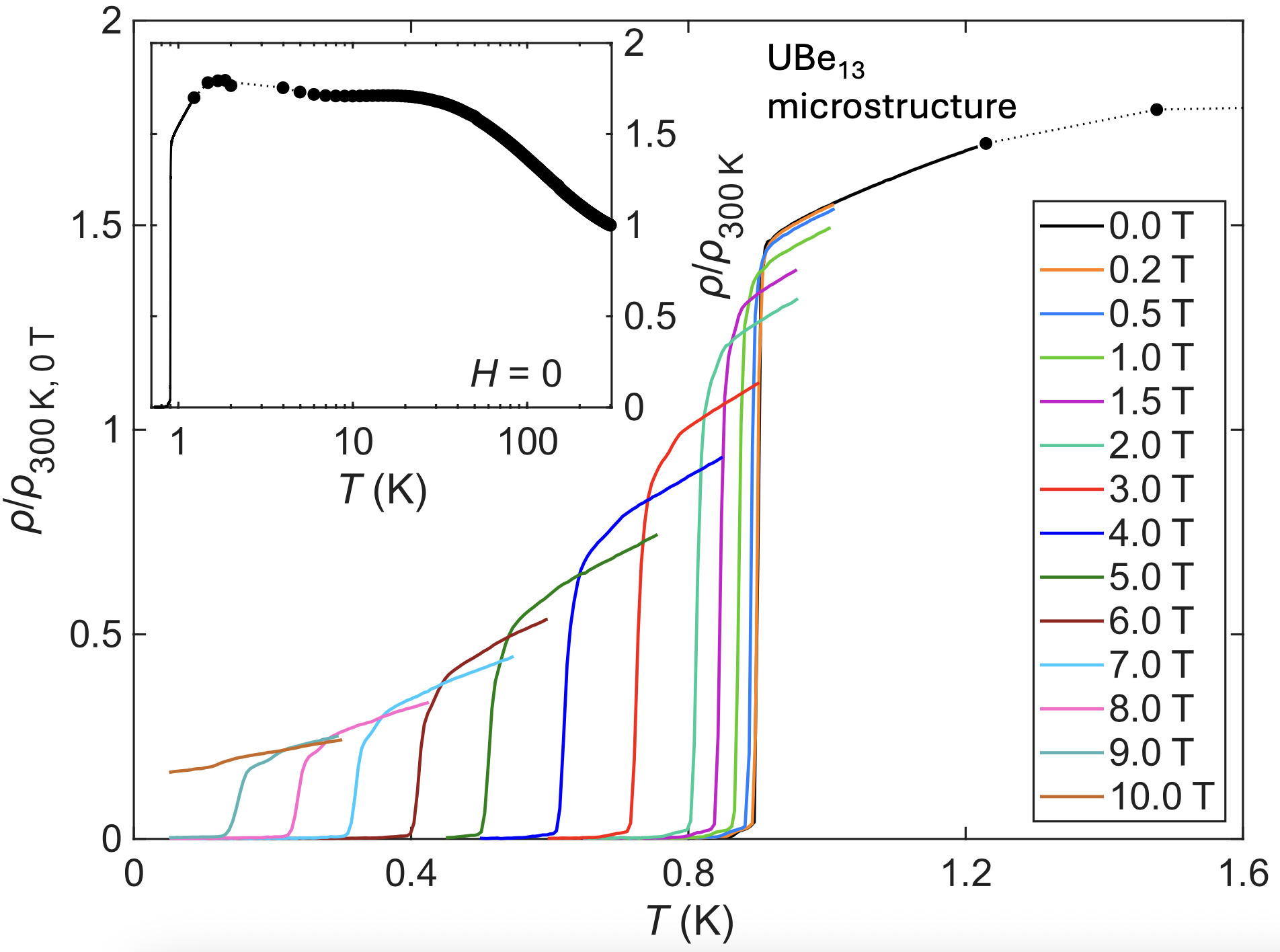}
\caption{Resistance as a function of temperature of a micro-scale UBe$_{13}$ device at different magnetic fields. $H_\mathrm{c2}$ data from these measurements are shown as red stars in Fig.~4 of the main paper.}
\label{fig:rhovsTmicro}
\end{figure}

First, the resistivity of three single crystals grown in aluminum flux (UBe$_{13-x}$Al$_x$) was measured in a dilution refrigerator down to $T = 60$ mK in magnetic fields up to $H = 15$ T. In Figs.~\ref{fig:rhovsTsc},~\ref{fig:rhovsHsc}, and ~\ref{fig:rhovsTmoredata}, we show exemplary resistivity data for two of them corresponding to upward and downward triangles in Fig.~4 of the main text. The sample with higher $T_c$ (labeled "annealed" in the figures) has a stronger coherence peak at around $T = 2$ K, a lower resistivity just above the superconducting state and a lower resistivity at $H = 15$\,T compared with the other sample. Here, the resistivity of the samples was normalized to $100\,\mu\Omega$cm at room temperature. Furthermore, only the annealed sample shows a a strong upturn in the $H_\mathrm{c2}$ curve.

Second, we have applied the focused-ion-beam isolation method \cite{Moll2018} to extract single crystallites of UBe$_{13-x}$ from polycrystalline material, which was previously shown to be exceptionally useful for multi-phase materials \cite{Amon2019,Antonyshyn2020,Gofryk2026}. 

Moreover, the bulk (from specific heat) and micro-scale (from micro-scale resistivity) superconducting critical temperatures of polycrystalline material (triangles vs. circles in Fig.~3(d) are in a good agreement with each other. This indicates that the composition of the micro-scale grains is homogeneous through the polycrystalline UBe$_{13-x}$ samples of this work.

AC electrical resistivity measurements were performed on a QD PPMS, using a standard four-probe technique at temperatures between $T = 0.4$ and $300$ K in $H = 0$. A current pulse of 0.01 mA with frequency 93 Hz for 1 s was applied along a piece of UBe$_{13-x}$. Voltage pairs were chosen so as to avoid U or Be inclusions (see Fig.~2). No evidence for anisotropy was found as multiple devices and voltage pairs were measured for a given sample -- isotropic properties are typically expected for a cubic system. Subsequent measurements in a dilution refrigerator are shown in Fig.~\ref{fig:rhovsTmicro} and used to determine the critical field curve shown as red stars in Fig.~4 of the main paper.

\begin{center}
\vspace{10pt}
\textbf{Strain induced by differential thermal expansion}
\end{center}

In Fig.~\ref{fig:differential thermal expansion}, we compare the thermal expansion of UBe$_{13}$ (from Ref. \cite{Hidaka2018}) with the one of pure Be and pure U in order to evaluate how much strain is induced in polycrystalline material by the Be (Ref. \cite{Bodryakov2018}) and U inclusions (Ref. \cite{Schuch1952}), respectively. The differential expansion between U and UBe$_{13}$ is larger than the one between Be and UBe$_{13}$. However, in either case, the maximum strain amounts to only -0.15~\%.
%%%%%%%%%%%%%%%%%%%%%%%%%%%%%%%%%%%%%%%%%%%%%%%%%%%%%%%%%%%%%%%%%%%%%%%%%%%%%%%%%%%%%

\end{document}